# Femtosecond Excitation Correlation Spectroscopy of Single-Walled Carbon Nanotubes: Analysis Based on Nonradiative Multiexciton Recombination Processes

Yuhei Miyauchi[1,*], Kazunari Matsuda[1], and Yoshihiko Kanemitsu[1,2]

[1]*Institute for Chemical Research, Kyoto University, Uji, Kyoto 611-0011, Japan*
[2]*Photonics and Electronics Science and Engineering Center, Kyoto University, Kyoto 615-8510, Japan*

## Abstract

We studied the nonlinear time-resolved luminescence signals due to multiexciton recombination processes in single-walled carbon nanotubes (SWNTs) using femtosecond excitation correlation (FEC) spectroscopy. From theoretical analysis of the FEC signals, we found that the FEC signals in the long time range are dominated by the single exciton decay in SWNTs, where the exciton–exciton annihilation process is efficient. Our results provide a simple method to clarify the single exciton decay dynamics in low-dimensional materials.

*Corresponding author

Electronic address: y.miyauchi@at7.ecs.kyoto-u.ac.jp

**I. Introduction**

Since the first report in 1981 [1, 2], the nonlinear time-resolved photoluminescence (PL) spectroscopy technique termed picosecond or femtosecond excitation correlation (PEC or FEC) has been applied to investigate the carrier and/or exciton dynamics in various semiconductors [1–15]. This method has the benefits of excellent time-resolution, limited only by the pulse width of the laser light used to excite the system, and a simpler experimental setup than the other ultrafast techniques. Theoretical models of the origin of FEC signals are inevitably required for interpretation of the data. Previous studies using the excitation correlation method concerned the recombination lifetimes of carriers [1–7] and excitons [8], tunneling dynamics [9–11], bimolecular formation of excitons [12, 13], and transport properties of carriers [14, 15]. However, no reports have addressed the FEC signals based on the optical nonlinearity originating from the exciton–exciton annihilation process.

The exciton–exciton annihilation process dominates the nonlinearity of low-dimensional semiconductor nanostructures with strong Coulomb interactions, such as single-walled carbon nanotubes (SWNTs) [16–20]. SWNTs have large exciton binding energies (~0.4 eV [21]), and excitons are stable even at room temperature. In addition, very rapid exciton–exciton annihilation processes (of the order 1 ps for only two excitons) have been reported [19, 20]. SWNTs show near IR PL around 1 eV, and the emission energy is inversely proportional to the tube diameter, *d*. Recent time-resolved measurements by time-correlated single photon counting (TCSPC) method [22–24], streak camera [25, 26], frequency up conversion [27], and Kerr gate [16, 28] techniques have revealed the PL lifetimes of isolated SWNTs of the order of 10–100 ps. However, the TCSPC measurement is inadequate for the large diameter SWNTs ($d > \sim 0.8$ nm) because of the limit of the sensitive range of Si-based single photon counting avalanche photodiodes, and the sensitivity and time-resolution of near IR streak cameras is low. Although frequency upconversion [27] and Kerr gate [16, 28] methods provide excellent time-resolution, these methods involve relatively complicated experimental setups. Hence, the development of alternatives has been eagerly anticipated to measure the ultrafast PL dynamics in SWNTs in the near IR range with good time-resolution.

In this paper, we calculate the nonlinear FEC signals arising from the nonradiative exciton–exciton annihilation process, and apply FEC spectroscopy to SWNTs. In Section II, we describe a theoretical analysis of the nonlinear FEC signals after introducing a general aspect of the FEC measurement. In Section III, we show the experimental procedures and results on SWNTs as a typical material having stable excitons even at room temperature and exhibiting very efficient exciton–exciton annihilation in the strong excitation regime. We analyze the experimental data based on the theoretical scheme presented in Section II and successfully derive the single exciton lifetimes in SWNTs. Our results suggest that the FEC signals are dominated by single exciton decay at longer delay times, after the rapid exciton–exciton nonradiative recombination processes. The single exciton lifetimes can be readily determined using the FEC technique in materials where the exciton–exciton annihilation process dominates the optical nonlinearity.

**II. Theoretical analysis of excitation correlation signals due to rapid exciton–exciton annihilation**

The PL correlation signals measured by the FEC method originate from the nonlinearity in the PL intensity as a function of the excitation power. We assume a pair of pump beams with the same power and photon energy, and the beams are modulated at two different frequencies, $\omega_1$ and $\omega_2$, and the nonlinear PL signal from the sample is detected using a lock-in amplifier. When one of the pump beams is blocked, the output signals of the lock-in amplifier are expressed as

$$I_1(t) = I_0 S(t, \omega_1), \qquad (1)$$

$$I_2(t) = I_0 S(t, \omega_2), \qquad (2)$$

where $I_0$ is the amplitude of the detected PL intensity, and $S(t, \omega_i)$ are 50% duty cycle square waves with amplitudes between 1 and 0 and frequencies $\omega_1$ and $\omega_2$. Nonlinear response that we detect as the FEC signals occurs only when the two pump beams are on at the same time. The time dependence of this signal can be expressed as

$$I_{\text{PLcor}}(t) = I_C(\tau) S(t, \omega_1) S(t, \omega_2), \qquad (3)$$

where $I_C(\tau)$ is the amplitude of the nonlinear correlation signal as a function of the delay time $\tau$ between two pump pulses. Then, the total detected signal can be expressed as

$$I_{\text{PLtot}}(t) = I_0 (S(t, \omega_1) + S(t, \omega_2)) + I_C(\tau) S(t, \omega_1) S(t, \omega_2). \qquad (4)$$

Because the term $S(t, \omega_1) S(t, \omega_2)$ can be decomposed into two components with frequencies $\omega_1 + \omega_2$ and $\omega_1 - \omega_2$, we can separate the contribution of $I_C(\tau)$ by selecting the lock-in response at the sum or difference frequencies. The functional form of $I_C(\tau)$ depends on the mechanism of nonlinearity in the materials.

Here we consider low dimensional semiconductors where the optical properties are dominated by strongly bound excitons, and the nonlinearity of the PL is mainly caused by exciton–exciton annihilation (Auger nonradiative recombination). This situation is well known to occur in SWNTs [16–18], which is one of the ideal one-dimensional materials with stable excitons and strong exciton–exciton interactions. Under strong pump intensity, where the initially generated exciton population in a SWNT by a single pump pulse is more than 1, we assume that the exciton population $N$ obeys the following rate equations:

$$\frac{dN_A}{dt} = G(t) - \gamma_1 N_A - \gamma_A N_A (N_A - 1) \quad (N_A \geq 1), \tag{5}$$

$$\frac{dN_1}{dt} = -\gamma_1 N_1 \quad (N_1 \leq 1), \tag{6}$$

where $N_A$ and $N_1$ are exciton populations under $N \geq 1$ and $N \leq 1$ conditions, respectively, $\gamma_A$ is the coefficient determining the nonlinear decay rate $\gamma_A N(N-1)$ [16], $\gamma_1$ is the recombination rate for a single exciton, and $G(t)$ is the instantaneous generation function of $N_0$ excitons at $t = 0$. Here we neglect the fine structure of the exciton levels (e.g., bright–dark exciton level splitting [29]), and assume that the excitons generated in higher energy states relax into the lowest bright state in a time shorter than the pulse duration, and unity relaxation efficiency. Under these conditions, Eqs. (5) and (6) become

$$N_A(t, N_0) = \frac{1}{\Gamma - (\Gamma - \frac{1}{N_0}) \exp(-(\gamma_A - \gamma_1)t)} \quad (N > 1), \tag{7}$$

$$N_1(t, t_1) = \exp(-\gamma_1 (t - t_1)) \quad (N \leq 1), \tag{8}$$

where $\Gamma \equiv \gamma_A / (\gamma_A - \gamma_1)$, and $t_1$ is the time when $N_A(t, N_0) = 1$. $t_1$ is expressed as

$$t_1 = \frac{1}{\gamma_A - \gamma_1} \ln\left(\frac{N_0 \Gamma - 1}{N_0 (\Gamma - 1)}\right). \tag{9}$$

For a single-pulse excitation, the exciton number $N_S(t, N_0)$ is expressed as

$$N_S(t, N_0) = (1 - \Theta(t - t_1)) N_A(t, N_0) + \Theta(t - t_1) N_1(t - t_1), \tag{10}$$

where $\Theta(t)$ is the heaviside step function for which $\Theta(0) = 1$ is defined. For a two-pulse excitation with a delay time $\tau$, the exciton number, $N_T(t, N_0, \tau)$, is given by

$$N_T(t, N_0, \tau) = \Theta(t)(1 - \Theta(t - \tau))N_S(t, N_0) + \Theta(t - \tau)N_S(t - \tau, N_0 + N_S(\tau, N_0)). \quad (11)$$

For recombination of the $N_T(t, N_0, \tau)$ excitons, the total PL intensity $I_{PL}(N_0, \tau)$ is described as

$$I_{PL}(N_0, \tau) = \int_0^\infty \gamma_R N_T(t, N_0, \tau) dt. \quad (12)$$

The correlation (FEC) signal $I_C(N_0, \tau)$ is defined as

$$I_C(N_0, \tau) = I_{PL}(N_0, \tau) - I_{PL}(N_0, \infty). \quad (13)$$

To evaluate $I_C(N_0, \tau)$, we calculate $I_{PL}(N_0, \tau)$ according to Eq. (12). Using $N_S(t, N_0)$, Eq. (12) becomes

$$I_{PL}(N_0, \tau) = \int_0^\tau \gamma_R N_S(t, N_0) dt + \int_\tau^\infty \gamma_R N_S(t - \tau, N_0 + N_S(\tau, N_0)) dt. \quad (14)$$

$I_{PL}(N_0, \tau)$ is calculated for $\tau < t_1$ and $\tau \geq t_1$ as

$$I_{PL}(N_0, \tau) = \frac{\gamma_R}{\gamma_A} \ln\left(\frac{[N_0 \Gamma \exp((\gamma_A - \gamma_1)\tau) - N_0 \Gamma + 1][(N_0 + N_A(\tau, N_0))\Gamma - 1]}{\Gamma - 1}\right) + \frac{\gamma_R}{\gamma_1}$$

($\tau < t_1$), (15)

$$I_{PL}(N_0, \tau) = \frac{\gamma_R}{\gamma_A} \ln\left(\frac{[N_0 \Gamma - 1][(N_0 + N_1(\tau, N_0))\Gamma - 1]}{(\Gamma - 1)^2}\right) + \frac{\gamma_R}{\gamma_1}[2 - \exp(-\gamma_1(\tau - t_1))]$$

($\tau \geq t_1$). (16)

$I_{PL}(N_0, \infty)$ is obtained by taking the limit of $\tau \to \infty$ as

$$I_{PL}(N_0, \infty) = 2\left[\frac{\gamma_R}{\gamma_A} \ln\left(\frac{N_0 \Gamma - 1}{\Gamma - 1}\right) + \frac{\gamma_R}{\gamma_1}\right]. \quad (17)$$

The $I_C(N_0, \tau)$ is thus calculated according to Eq. (13) as

$$I_C(N_0,\tau) = -\frac{\gamma_R}{\gamma_1} + \frac{\gamma_R}{\gamma_A}\ln\left(\frac{[N_0\Gamma\exp((\gamma_A-\gamma_1)\tau)-N_0\Gamma+1][(N_0+N_A(\tau,N_0))\Gamma-1](\Gamma-1)}{(N_0\Gamma-1)^2}\right)$$

$(\tau < t_1),$ (18)

$$I_C(N_0,\tau) = -\frac{\gamma_R}{\gamma_1}\exp(-\gamma_1(\tau-t_1)) + \frac{\gamma_R}{\gamma_A}\ln\left(\frac{(N_0+N_1(\tau,t_1))\Gamma-1}{N_0\Gamma-1}\right)$$

$(\tau \geq t_1).$ (19)

Note that the second logarithm term in Eq. (19) is considerably smaller than the first exponential term for $N_0 \gg 1$ and $\gamma_A \gg \gamma_1$. The simplified form of $I_C(N_0,\tau)$ in this condition is therefore

$$I_C(N_0,\tau) = -\frac{\gamma_R}{\gamma_1}\left(\frac{N_0\Gamma-1}{N_0(\Gamma-1)}\right)^{\frac{\gamma_1}{\gamma_A-\gamma_1}}\exp(-\gamma_1\tau).$$ (20)

Hence, the correlation signals for $\tau \geq t_1$, $N_0 \gg 1$ and $\gamma_A \gg \gamma_1$ can be approximated as simple mono-exponential decay with the total exciton recombination rate $\gamma_1$.

The theoretical result can be qualitatively understood as follows. Figures 1(a) and 1(b) show schematic diagrams of the mechanism of FEC signals due to exciton–exciton annihilation for $\tau \gg \gamma_1^{-1}$ and $\tau \sim \gamma_1^{-1}$ under the conditions $N_0 \gg 1$ and $\gamma_A \gg \gamma_1$, respectively. The first pulse generates $N_0$ excitons at $t = 0$, and the multiexcitons quickly annihilate due to rapid exciton–exciton nonradiative recombination processes, followed by the decay of the single surviving exciton. At longer delay time $\tau \gg \gamma_1^{-1}$ in Fig. 1(a), no correlation exists between the PL signals generated by the first and second pulse; the nonlinear correlation signal is zero. After the shorter delay time, $\tau \sim \gamma_1^{-1}$, in Fig. 1(b), the second pulse additionally generates $N_0$ excitons, and the total generated, $N_0 + N_1(\tau)$, ($N_1(\tau) < 1$) again quickly annihilate. If we neglect the integrated PL intensity in the very short time period of the fast nonlinear decay ($0 \leq \tau \leq t_1$), the second pulse instantaneously removes the $N_1(\tau)$ excitons generated by the first pulse. The difference of $I_{PL}(N_0, \tau)$ and $I_{PL}(N_0, \infty)$ thus corresponds to the integration of the removed excitons from $t = \tau$ to $\infty$ as

$$I_C(N_0, \tau) \approx -\int_\tau^\infty \gamma_R N_1(t) dt. \qquad (21)$$

Hence, if $N_1(t)$ is the mono-exponential function, the correlation signal can be approximated by a mono-exponential function in this case.

Figures 2(a) and 2(b) show simulated correlation signals using Eqs. (18) and (19) for various $N_0$ values with the parameters of $\gamma_A / \gamma_1 = 10$ and $\gamma_A / \gamma_1 = 100$, respectively. These values were selected for simulation because the reported nonlinear annihilation coefficients $\gamma_A$ are of the order of 1 ps$^{-1}$ [19, 20] for SWNTs, which is about one to two orders of magnitude larger than the single exciton recombination rates [22–28]. The upward direction on the vertical axis in the figures indicates that the FEC signals have a negative sign. A mono-exponential decay is also shown in Figs. 2(a) and 2(b) for comparison. The weak dependence on $N_0$ appears only in the very short delay time range, [Figs. 2(a) and 2(b)] when the single exciton recombination rate is much smaller than the exciton–exciton annihilation rate. After a delay time longer than $\sim \gamma_A^{-1}$, the decay curve is almost perfectly coincident with the mono-exponential decay with the single-exciton decay rate $\gamma_1$. This indicates that the single exciton decay can be readily studied from the FEC signals as long as the condition $\gamma_A \gg \gamma_1$ is satisfied. Moreover, we can check whether these conditions are satisfied by measuring the excitation power dependence of the FEC decay curves.

**III. Experimental results of femtosecond excitation correlation signals from single-walled carbon nanotubes**

The SWNTs were synthesized by the alcohol catalytic chemical vapor deposition (CCVD) method at 850°C [30]. These SWNTs were isolated by dispersion in a toluene solution with 0.07 wt% poly[9,9-dioctylfluorenyl-2,7-diyl] (PFO) (PFO-dispersed SWNTs), 60 minutes of moderate bath sonication, 15 minutes of vigorous sonication with a tip-type sonicator, and centrifugation at an acceleration of 13 000 g for 5 min, according to the procedure developed by Nish et al. [31].

We measured the delay time dependence of FEC signals for SWNTs. The SWNTs were excited with ultra-short pulses from a Ti:sapphire laser of central wavelength 745 nm, repetition rate 80 MHz, pulse duration ~150 fs, and spectral width 8 nm. The two beams were separated by a delay time $\tau$ and chopped at 800 and 670 Hz, respectively, then collinearly focused to a spot size of ~10 μm. Only the PL signal components modulated at the sum frequency (1470 Hz) were detected using a photomultiplier and a lock-in amplifier, following dispersion of the PL using a monochromator. The measurements were carried out under the excitation of ~20 to 300 μJ/cm$^2$. The background subtraction of the FEC signals was based on the signal values at delay times more than ~0.7-1 ns, which is considerably longer than the PL lifetimes of SWNTs. We have confirmed the validity of the background subtraction from the experimental observation that no PL signal in the range of more than ~600 ps exists using a near IR streak camera with the time resolution of ~100 ps.

Figures 3(a) and 3(b) show an optical absorption spectrum and a PL excitation map of PFO-dispersed SWNTs, respectively. The very low underlying background in the absorption spectrum and pronounced absorption and PL peaks are a signature of excellently isolated, high quality dispersion of SWNTs with an absence of bundled SWNTs, residual impurities, or other amorphous or graphitic carbon compounds [31]. The optical measurements indicate that only several types of chiral indices ($n$, $m$) [32] are included in the sample.

Figure 4 shows the excitation power dependence of the FEC signals as a function of the delay time for (8, 7) SWNTs, normalized at $\tau = 0$. We observed no change of the FEC decay curve with the excitation power density in the range ~20 to 300 µJ/cm$^2$, which is consistent with previous results for the micelle-encapsulated SWNTs in D$_2$O [33]. Assuming the recently reported absorption cross section of $E_{22}$ excitons ~110 nm$^2$/µm [24], the number of excitons generated by a single pulse in our experiment was roughly estimated as ~10$^1$–10$^2$ excitons for the excitation density of ~20 to 300 µJ/cm$^2$. From this estimate, we can confirm that the theoretical analysis in Sec. II is applicable to the experimentally obtained FEC signals. Moreover, this lack of excitation power dependence is a strong indication of the very rapid exciton–exciton annihilation processes in SWNTs described in Sec. II.

Figure 5 shows the FEC signals for various nanotube species. We found that the decay curve is well described by a double-exponential function (solid line) after subtracting the background signals for all the observed ($n$, $m$) SWNTs. The double-exponential PL decay of a single exciton in SWNTs was recently observed using single nanotube spectroscopy [24]. Since we observed no excitation power dependence of the FEC decay curve (as shown in Fig. 4), we can use Eq. (21) for the analysis of the FEC signals. The exciton population obeying the double-exponential decay as

$$N_1(t) = C\exp(-t/\tau_A) + (1-C)\exp(-t/\tau_B) \quad (0 \leq C \leq 1, \tau_A < \tau_B) \quad (22)$$

gives the FEC signal $I_C(\tau)$ calculated using Eq. (21) as

$$I_C(\tau) \propto -\left[C\tau_A \exp(-\tau/\tau_A) + (1-C)\tau_B \exp(-\tau/\tau_B)\right], \quad (23)$$

where $C$ is the fractional amplitude of the fast decay component. For (7, 5) SWNTs, we fitted the experimental results of FEC signals and obtained $C \cong 0.92$, $\tau_A \cong 45$ ps and $\tau_B \cong 200$ ps. Here we define the effective PL lifetime as $\tau_{EFF} = C\tau_A + (1-C)\tau_B$. The $\tau_{EFF}$ obtained for (7, 5) was $58 \pm 21$ ps. We have also checked that $\tau_{EFF}$ is consistent with that obtained by the streak camera (~ 60 ps). Furthermore, these results are quite similar to the recently reported values for single (6, 5) SWNTs in surfactant suspension measured by TCSPC [24]. This suggests that the single exciton decay can be properly measured by the FEC technique for SWNTs.

We also measured the PL lifetimes of various (*n*, *m*) species, as shown in Fig. 5. The $\tau_{EFF}$ values obtained for (7, 6), (8, 6), and (8, 7) SWNTs were $43 \pm 9$, $42 \pm 12$, and $33 \pm 9$ ps, respectively. The larger diameter SWNTs tend to have shorter PL lifetimes. Because the radiative lifetimes of excitons in SWNTs are of the order of 1–10 ns [34], the PL lifetimes of the order of several tens of picoseconds are attributed to non-radiative decay due to extrinsic effects such as defects and impurities. Because the larger diameter SWNTs (*d* > 1nm) are important for applications such as optical communication devices, suppression of the non-radiative decay and improvement of the PL quantum yields for larger diameter SWNTs is highly desirable. In addition to the usefulness of the FEC technique for fundamental physics research, PL lifetime measurement using FEC will enable easy sample-quality screening of large diameter SWNTs with PL emission energies less than ~1 eV.

**IV. Summary**

We have demonstrated the theoretical analysis of FEC signals originating from nonradiative exciton–exciton annihilation processes. We found that the FEC signals are dominated by the single exciton decay dynamics when the nonlinear exciton–exciton annihilation process is much faster than the single exciton decay. We measured the FEC signals in SWNTs and determined the single-exciton decay lifetimes. Our results suggest that FEC spectroscopy, with advantages in terms of the time-resolution and simple experimental setup, is very useful for the exciton lifetime measurement in low-dimensional materials with rapid exciton–exciton annihilation.

**Acknowledgements**

The authors would like to thank Prof. S. Noda and Mr. K. Ishizaki (Kyoto University) for their experimental support in using the IR streak camera. One of the authors was supported by a JSPS Research Fellowships for Young Scientists Award. Part of this work was supported by JSPS KAKENHI (No. 20340075) and MEXT KAKENHI (Nos. 20048004 and 20104006) of Japan.

# References

[1] D. Rosen, A. G. Doukas, Y. Budansky, A. Katz, and R. R. Alfano, Appl. Phys. Lett. **39**, 935 (1981).

[2] D. von der Linde, J. Kuhl, and E. Rosengart, J. Lumin. **24/25**, 675 (1981).

[3] M. B. Johnson, T. C. McGill, and A. T. Hunter, J. Appl. Phys. **63**, 2077 (1988).

[4] H. J. W. Eakin and J. F. Ryan, J. Lumin. **40/41**, 553 (1988).

[5] A. M. de Paula, R. A. Taylor, C. W. W. Bradley, A. J. Turberfield, and J. F. Ryan, Superlattices Microstruct. **6,** 199 (1989).

[6] J. L. A. Chilla, O. Buccafusca, and J. J. Rocca, Phys. Rev. B **48**, 14347 (1993).

[7] R. Kumar and A. S. Vengurlekar, Phys. Rev. B **54**, 10292 (1996).

[8] M. Jørgensen and J. M. Hvam, Appl. Phys. Lett. **43**, 460 (1983).

[9] M. K. Jackson, M. B. Johnson, D. H. Chow, T. C. McGill, and C. W. Nieh, Appl. Phys. Lett. **54**, 552 (1989).

[10] N. Sawaki, R. A. Höpfel, E. Gornik, and H. Kano, Appl. Phys. Lett. **55**, 1996 (1989).

[11] V. Emiliani, S. Ceccherini, F. Bogani, M. Colocci, A. Frova, and S. S. Shi, Phys. Rev. B **56**, 4807 (1997).

[12] R. Strobel, R. Eccleston, J. Kuhl, and K. Köhler, Phys. Rev. B **43**, 12564 (1991).

[13] S. Pau, J. Kuhl, M. A. Khan, and C. J. Sun, Phys. Rev. B **58**, 12916 (1998).

[14] R. Christanell and R. A. Höpfel, Superlattices Microstruct. **5,** 193 (1989).

[15] E. Okuno, T. Hori, N. Sawaki, I. Akasaki, and R. A. Höpfel, Jpn. J. Appl. Phys. **31**, L148 (1992).

[16] F. Wang, G. Dukovic, E. Knoesel, L. E. Brus and T. F. Heinz, Phys. Rev. B **70**, 241403(R) (2004).

[17] Y.-Z. Ma, L. Valkunas, S. L. Dexheimer, S. M. Bachilo, and G. R. Fleming, Phys. Rev.

Lett. **94**, 157402 (2005).

[18] L. Valkunas, Y.-Z. Ma, and G. R. Fleming, Phys. Rev. B **73**, 115432 (2006).

[19] F. Wang, Y. Wu, M. S. Hybertsen, and T. F. Heinz, Phys. Rev. B **73**, 245424 (2006).

[20] K. Matsuda, T. Inoue, Y. Murakami, S. Maruyama, and Y. Kanemitsu, Phys. Rev. B **77**, 033406 (2008).

[21] F. Wang, G. Dukovic, L. E. Brus, and T. F. Heinz, Science **308,** 838 (2005).

[22] A. Hagen, M. Steiner, M. B. Raschke, C. Lienau, T. Hertel, H. Qian, A. J. Meixner, and A. Hartschuh, Phys. Rev. Lett. **95**, 197401 (2005).

[23] M. Jones, W. K. Metzger, T. J. McDonald, C. Engtrakul, R. J. Ellingson, G. Rumbles, and M. J. Heben, Nano Lett. **7,** 300 (2007).

[24] S. Berciaud, L. Cognet, and B. Lounis, Phys. Rev. Lett. **101**, 077402 (2008).

[25] A. Hagen, G. Moos, V. Talalaev, and T. Hertel, Appl. Phys. A. **78**, 1137-1145 (2004).

[26] S. Berger, C. Voisin, G. Cassabois, C. Delalande, P. Roussignol, and X. Marie, Nano Lett. **7,** 398 (2007).

[27] Y.-Z. Ma, J. Stenger, J. Zimmermann, S. M. Bachilo, R. E. Smalley, R. B. Weisman, and G. R. Fleming, J. Chem. Phys. **120**, 3368 (2004).

[28] F. Wang, G. Dukovic, L. E. Brus, and T. F. Heinz, Phys. Rev. Lett. **92** 177401 (2004).

[29] H. Zhao and S. Mazumdar, Phys. Rev. Lett. **93**, 157402 (2004).

[30] S. Maruyama, R. Kojima, Y. Miyauchi, S. Chiashi, and M. Kohno, Chem. Phys. Lett. **360,** 229 (2002).

[31] A. Nish, J.-Y Hwang, J. Doig, and R. J. Nicholas, Nat. Nanotechnol. **2,** 640 (2007).

[32] R. Saito, G. Dresselhaus, and M.S. Dresselhaus, *Physical Properties of Carbon Nanotubes*, Imperial College Press, London, 1998.

[33] H. Hirori, K. Matsuda, Y. Miyauchi, S. Maruyama, and Y. Kanemitsu, Phys. Rev.

Lett. **97,** 257401 (2006).

[34] Y. Miyauchi, H. Hirori, K. Matsuda, and Y. Kanemitsu, submitted for publication.

# Figure captions

**Figure 1.** (color online) Schematic diagram of the mechanism of FEC signals due to exciton–exciton annihilation for (a) $\tau \gg \gamma_1^{-1}$ and (b) $\tau \sim \gamma_1^{-1}$ under the conditions of $N_0 \gg 1$ and $\gamma_A \gg \gamma_1$. The shaded area in (b) corresponds to the FEC signal at decay time $\tau$.

**Figure 2.** (color online) Simulated FEC signals for various $N_0$ for (a) $\gamma_A / \gamma_1 = 10$ and (b) $\gamma_A / \gamma_1 = 100$. Mono-exponential FEC decays are plotted for comparison. Insets show the FEC signals in the long delay time range.

**Figure. 3.** (color online) (a) Optical absorption spectra of PFO-dispersed SWNTs at room temperature. (b) PL intensity map as a function of excitation and emission photon energies for PFO-dispersed SWNTs.

**Figure. 4.** (color online) (a) Fast-decay component of the FEC signals for the (7, 5) SWNTs of the PFO-dispersed sample measured under 1.66-eV excitation conditions from ~20 to 300 μJ/cm². The solid curve shows the fitted result using a mono-exponential function.

**Figure 5.** (color online) FEC signals for (7, 5), (7, 6), (8, 6) and (8, 7) SWNTs, respectively. The fitted curves using the double-exponential function in Eq. (23) are shown as solid lines.

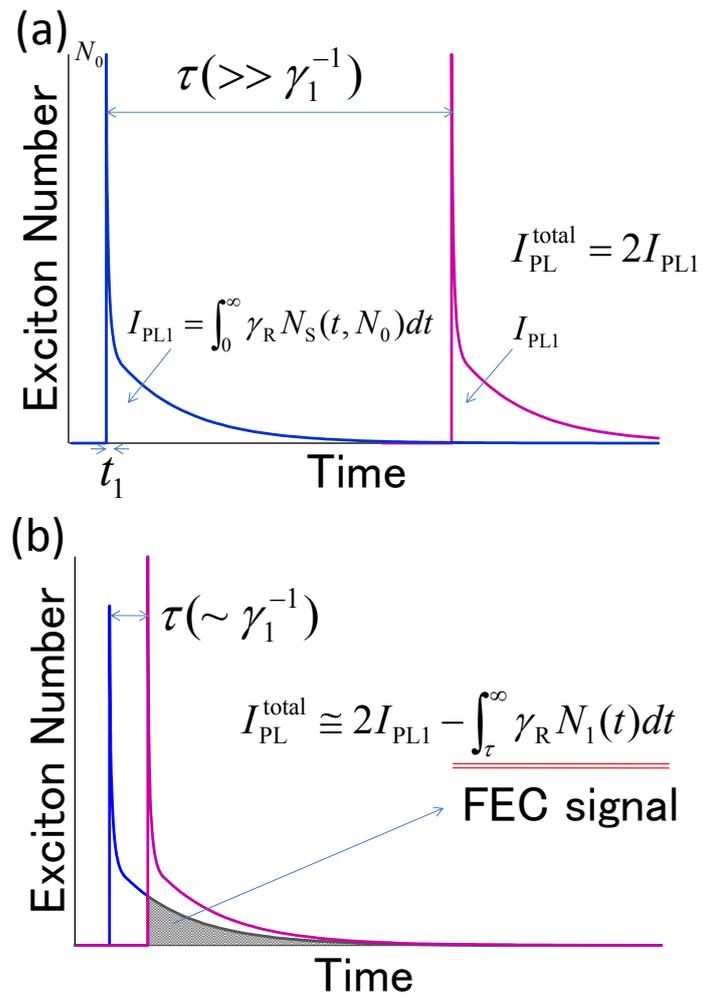

Fig.1 (color online) Y. Miyauchi et al.

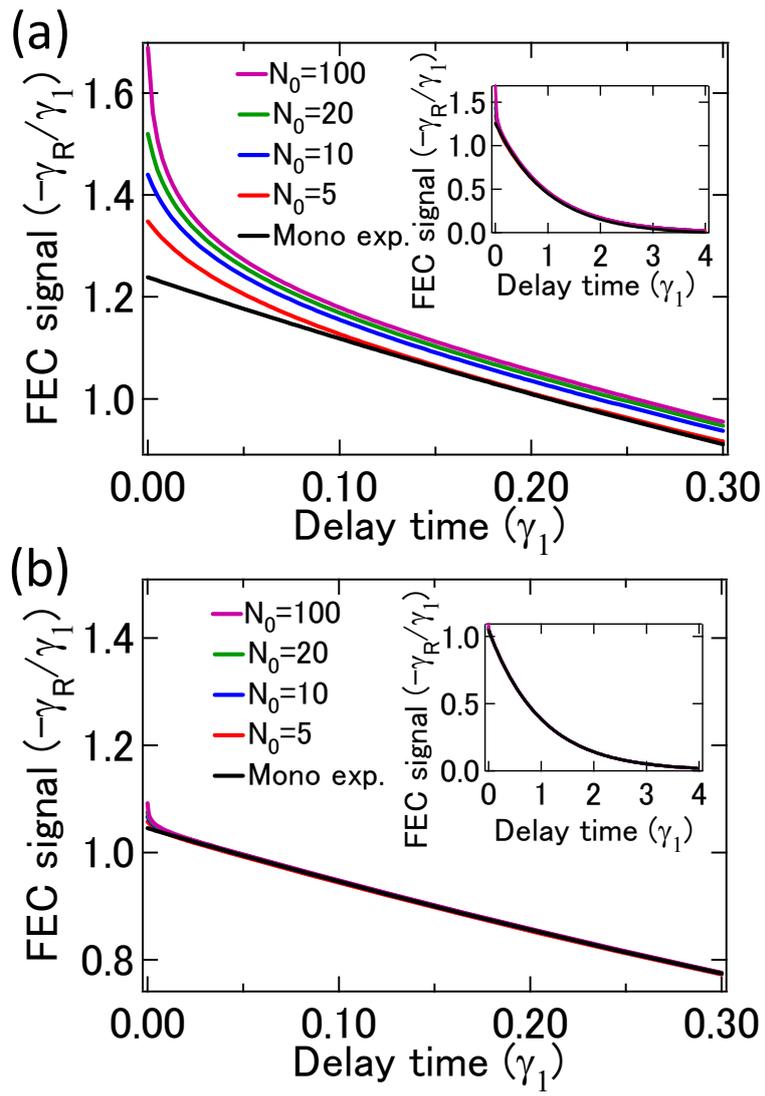

Fig. 2 (color online) Y. Miyauchi et al.

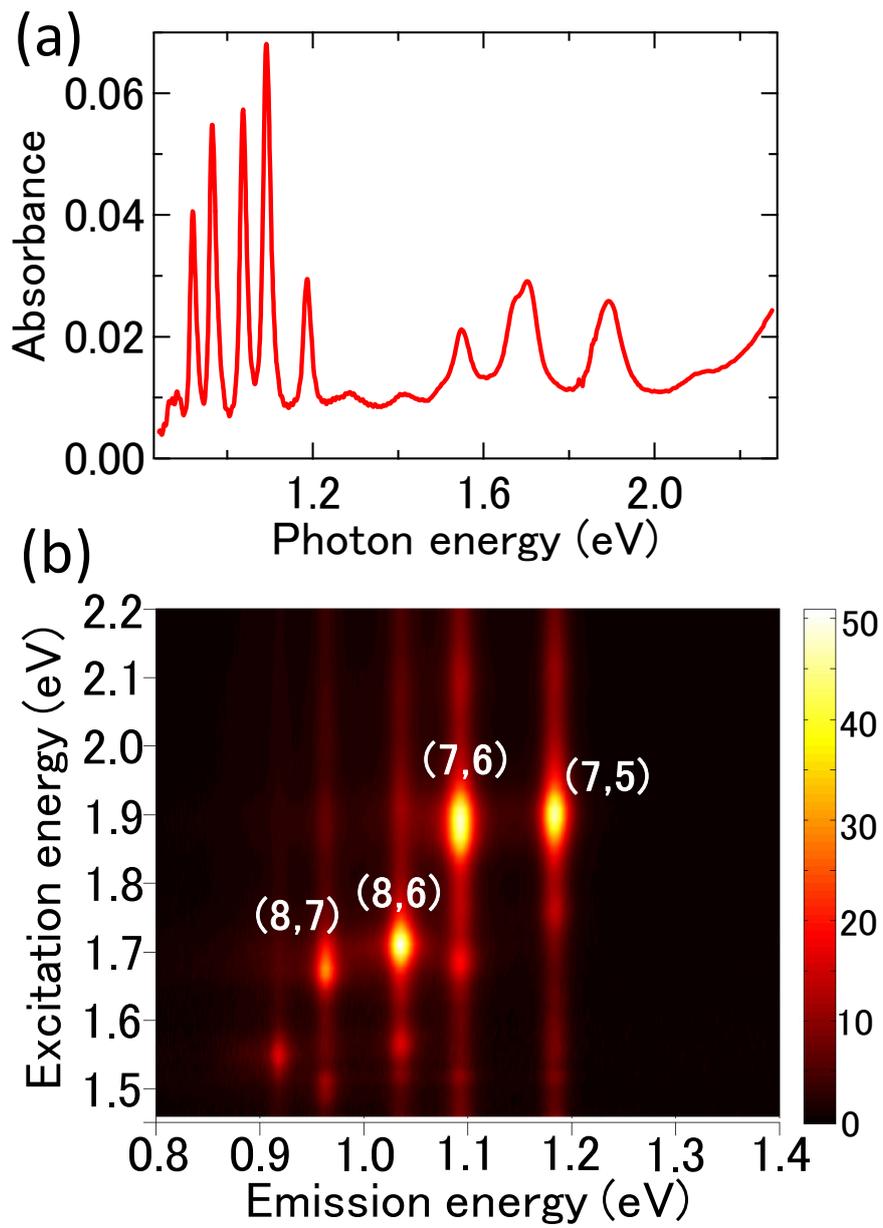

Fig. 3 (color online) Y. Miyauchi et al.

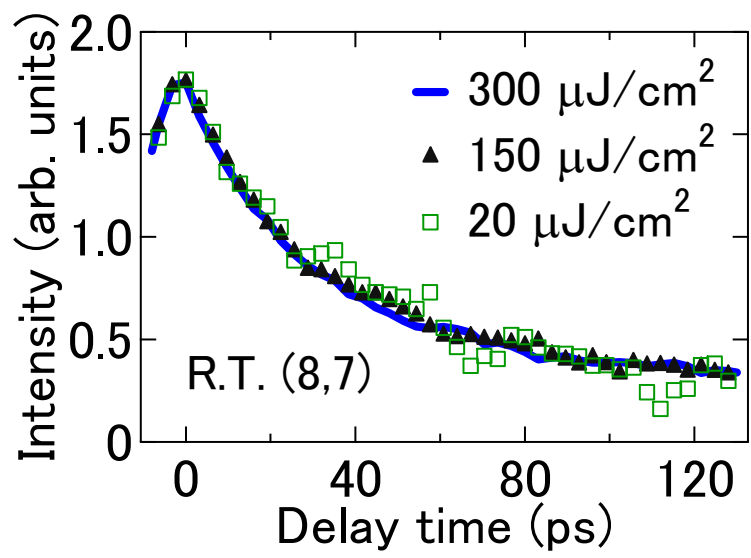

Fig. 4 (color online) Y. Miyauchi et al.

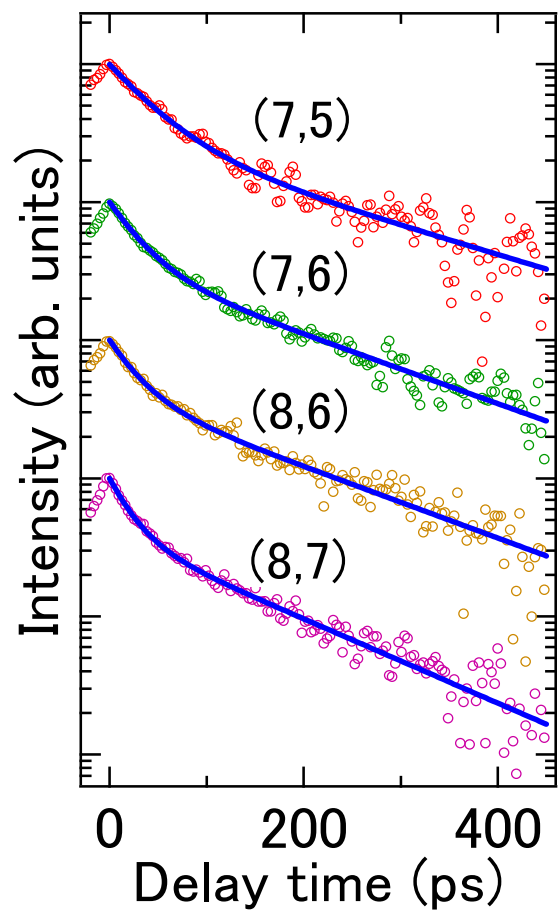

Fig. 5 (color online) Y. Miyauchi et al.